\documentclass[%
 reprint,
 amsmath,amssymb,
prl,
]{revtex4-2}

\usepackage{graphicx}
\usepackage{dcolumn}
\usepackage{bm}
\usepackage{braket}
\usepackage{stmaryrd}
\usepackage{color}
\usepackage{hyperref}
\usepackage{booktabs}
\usepackage{tikz}
\usepackage{tikz-cd}
\usepackage{multirow}
\usetikzlibrary{quantikz2}

\makeatletter

\newcommand*{\wideboxed}[1]{\setlength{\fboxsep}{1ex}%
  \fbox{\m@th$\displaystyle#1$}}
\makeatother

\begin{document}

\preprint{APS/123-QED}

\title{Magic teleportation with generalized lattice surgery}

\author{Yifei Wang}
\author{Yingfei Gu}%
 \email{guyingfei@tsinghua.edu.cn}
\affiliation{
  Institute for Advanced Study, Tsinghua University, Beijing 100084, China
}%

\date{\today}

\begin{abstract}
We propose a novel, distillation‐free scheme for the fault‐tolerant implementation of non-Clifford gates at the logical level, thereby completing the universal gate set. 
Our approach exploits generalized lattice surgery to integrate two quantum error-correcting (QEC) codes. Specifically, non-Clifford gates are executed transversally on one QEC code and then teleported to the main circuit via a logical-level joint measurement that connects two distinct QEC codes. In contrast to conventional magic state distillation (MSD) combined with gate teleportation, our method obviates the need for concatenating separate codes for distillation and logical qubits, thus reducing the total overhead from multiplicative to additive scaling. 
We illustrate our approach by explicitly demonstrating its implementation for a 3D color code interfaced with a surface code of the same code distance,
and comment on its potential advantage over the conventional MSD-teleportation scheme. 
\end{abstract}

\maketitle

\emph{Introduction.} 
Large-scale quantum computation requires quantum error correction (QEC) to achieve the desired logical error rate~\cite{Shor_1995}.  
However, a single QEC code is insufficient to protect a universal set of quantum gates~\cite{Eastin_2009}. 
Fault tolerance for Clifford operations, including Clifford gates, Pauli measurements and state preparation in the computational basis, are relatively easy to achieve in stabilizer codes~\cite{Calderbank_1996, Steane_1996, gottesman1997stabilizercodesquantumerror}. 
In contrast, the fault tolerance of non-Clifford gates, which are indispensable for achieving computational power beyond classical computers 
require additional efforts~\cite{gottesman1998heisenberg, Campbell_2017}. 

The leading approach to complete the universal set of gates at logical level
and achieve fault-tolerant quantum computation (FTQC) 
is to 
consume so called ``magic states'' to teleport non-Clifford gates into the circuit run on logical qubit~\cite{Gottesman_1999}. 
These magic states at desired precision are produced through ``magic state distillation'' (MSD)~\cite{Bravyi_2005}, 
a technique that refines high-quality magic states from a larger number of noisy magic states. 
The overhead 
of MSD is quantified by the number of initial noisy magic states needed to yield one magic state with a specified error rate $\epsilon$, scaling as $\log^\gamma \epsilon^{-1}$ \cite{Bravyi_2005, Bravyi_2012}, assuming  perfect Clifford operations. 
Since the invention of MSD, the parameter $\gamma$ has been gradually reduced from values greater than two to zero \cite{Bravyi_2005, Bravyi_2012, Hastings_2018, Krishna_2019, wills_2024, golowich_2024, nguyen_2024}.

Despite these advances, the conventional distillation-teleportation scheme remains resource-intensive as the whole procedure is performed at logical level: 
the logical qubits involved in MSD 
and gate teleportation 
are protected by the base code that is used in the main circuit/logical processor that implements all the Clifford operations. 
This concatenated structure results in the total overhead
of the full fault-tolerant universal computation 
being the product of the overhead of MSD and the overhead of fault-tolerant Clifford gates.

In this paper,
we propose a novel distillation-free scheme to implement non-Clifford gates, where the total overhead is a ``sum'' of the overhead of the code for transversal non-Clifford and the code for Clifford gates, 
compared with the ``product'' structure in the distillation-teleportation scheme aforementioned. 
The key component in our scheme is a ``generalized gate teleportation'' that allows a non-Clifford gate on one QEC code to be teleported to the circuit run on a different QEC code. 
Our scheme is based on two threads of recent developments: 
(1) 
A series of QEC codes with transversal non-Clifford gates, including good codes~\cite{wills_2024, golowich_2024, nguyen_2024} and quantum low-density parity-check (qLDPC) codes~\cite{lin_2024, golowich_2024_ldpc, breuckmann_2024}, have been constructed. 
These codes make it possible to produce magic states with arbitrary precision directly from transversal physical operations. 
(2) The lattice surgery technique for Pauli measurements, which was first proposed for geometrically local logical operations on surface codes \cite{Horsman_2012}, has been generalized to arbitrary CSS codes, especially qLDPC codes \cite{Cohen_2022, cross_2024, ide_2024, swaroop_2024}. 
The generalized lattice surgery allows for the fault-tolerant joint measurement of logical Pauli operators between different qLDPC codes.

\emph{Generalized gate teleportation.} 
The standard gate teleportation protocol for implementing an $n$-qubit diagonal gate $U$ at logical level involves a resource state $\ket{\overline{M}} = \overline{U}\ket{\overline{+}}^{\otimes n}$,
Clifford operations for teleporting the information,
and corrections conditioned on the measurement outcomes \cite{Gottesman_1999}.
Take gate $T = \operatorname{diag}(1,\exp(\mathrm{i}\pi/4))$ for example.
The standard gate teleportation for $T$
is accomplished by the following (logical-level) circuit,
\begin{equation}
    \begin{quantikz}[align equals at=1.5, row sep = {0.7cm,between origins}]
    \lstick{$\ket{\overline{\psi}}$} & \ctrl{1} & & \gate{\overline{S}^m} & \rstick{$\overline{T}\ket{\overline{\psi}}$} \\
    \lstick{$\ket{\overline{M}}$} & \targ{} & \gate{M_{\overline{Z}}}\wire[r][1]["m"{above,pos=0.2}]{c} & \ctrl[vertical wire = c]{-1} \setwiretype{n}
  \end{quantikz},
  \label{eq: gt-traditional}
\end{equation}
where the state 
$|\overline{M}\rangle=\overline{T}\ket{\overline{+}}$ is the resource to consume, referred as ``magic state''. 
In the above circuit, the Clifford correction $\overline{S}$ is performed conditioned on the measurement outcome $m$ (i.e. $m=0$, $\overline{S}=I$; $m=1$, $\overline{S}^m=\overline{S}$), and the input qubit evolves to $\overline{T}\ket{\overline{\psi}}$. 
In this standard gate teleportation scheme, the whole circuit is naturally encoded in the same QEC code, referred to as the base code, that supports all Clifford operations.
The MSD is required to produce magic state $|\overline{M}\rangle$ encoded in the base code that matches the target fidelity \footnote{It may look different from what one usually sees in literature, 
where the output state $\overline{T}\ket{\overline{\psi}}$ is hosted by the ancillary qubit.
In such conventional gate teleportation schemes, 
whether the output is hosted by the same qubits as the input makes little difference to later computation procedures,
since both the input qubit and ancilla are protected by the same base code.
}.

Here we propose a generalized gate teleportation protocol that allows a distillation-free scheme to implement non-Clifford gates fault tolerantly.
In this protocol,
\begin{enumerate}
    \item the main circuit that hosts the input and output state 
    is encoded in a base code B(ase),
    while the ancillae that supply non-Clifford resources are encoded in another code M(agic);
    \item gate teleportation is performed 
    by joint measurement of Pauli operators between logical qubits in code B and M, and Pauli measurement on logical qubits in code M, together with corresponding corrections conditioned on the measurement outcomes on the main circuit (in code B).
\end{enumerate}
The flexibility of choosing two different codes for the main and ancilla logical qubits opens a door to more efficient FTQC scheme.

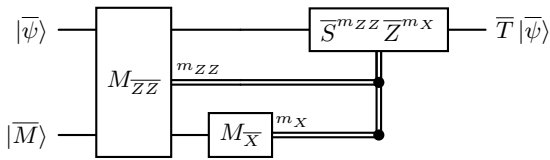
\begin{figure}[t]
    \centering
    \begin{quantikz}[align equals at=2, row sep = {0.7cm,between origins}]
    \lstick{$\ket{\overline{\psi}}$} & \gate[3]{{M}_{\overline{ZZ}}} & & \gate{\overline{S}^{m_{ZZ}}\overline{Z}^{m_X}} & \rstick{$\overline{T}\ket{\overline{\psi}}$}\\
    \setwiretype{n} & \wire[r][1]["m_{ZZ}"{above,pos=0.4}]{c} &  & \ctrl[vertical wire = c]{-1}\setwiretype{c} &\setwiretype{n}  \\ 
    \lstick{$\ket{\overline{M}}$} & & \gate{M_{\overline{X}}} \wire[r][1]["m_X"{above,pos=0.2}]{c}&\ctrl[vertical wire = c]{-2} \setwiretype{n}
  \end{quantikz}
    \caption{Teleportation of $T$ gate via lattice surgery between quibits encoded in different QEC codes. 
    The upper line started with $|\overline{\psi} \rangle$  represents the main circuit encoded in B, 
while the lower line represents the magic ancilla encoded in M.
In this case, $|\overline{M}\rangle =\overline{T}\ket{\overline{+}}$ is a the same logical state as the resource state in (\ref{eq: gt-traditional}), but is produced directly by applying transversal gate $\overline{T}$ to the logical state $\ket{\overline{+}}$.
After the joint measurement of $\overline{Z}\overline{Z}$ and $\overline{X}$, one need to apply the corrections conditioned on the outcomes $C(m_{ZZ},m_{X}) = \overline{S}^{m_{ZZ}}\overline{Z}^{m_X}$, which consists of only Clifford gates, explicitly demonstrated in Tab.~\ref{tab:cc}. 
This specific circuit also appeared in literature like \cite{litinskiGameSurfaceCodes2019}
for gate teleportation between surface code blocks.
}
    \label{fig:circuit}
\end{figure}

Specific to $T$-gate, using this new protocol, one can choose a large code M equipped with a transversal $T$  to generate $T$ state $|\overline{M}\rangle =\overline{T}\ket{\overline{+}}$ at arbitrary precision (for instance the 3D color code with large code distance, which will be discussed in detail momentarily). Then, the generalized gate teleportation 
can be implemented by the circuit demonstrated in Fig.~\ref{fig:circuit},
providing a direct channel to inject such magic resource to the main circuit. 
In this case, the quantum state $|\overline{\psi} \rangle$ under processing is encoded in certain base code B that supports all the Clifford operations. Then the $T$ gate is teleported by joint logical $\overline{ZZ}$ measurement between $|\overline{\psi} \rangle$ and $|\overline{M} \rangle$ that are encoded in different QEC codes, followed by a logical $\overline{X}$ measurement on ancilla qubits and a Clifford correction $C(m_{ZZ},m_{X}) = \overline{S}^{m_{ZZ}}\overline{Z}^{m_X}$ (also see Tab.~\ref{tab:cc}) on the main circuit.

Similar circuit can be constructed for multi-qubit non-Clifford diagonal gates such as C$S$ and CC$Z$ \footnote{See Supplemental Material A for an explicit construction and details on teleportation of a general diagonal gate.}.

\begin{table}
\centering
    \begin{tabular}{ccc}
    \toprule  
   $m_{ZZ}$ & $m_X$ & $C(m_{ZZ},m_X)$\\
   \midrule  
   $0$ & $0$ &  $I $ \\
    $0$ & $1$ &  $\overline{Z}$ \\
      $1$ & $0$ &  $\overline{S}$ \\
      $1$ & $1$ &  $\overline{SZ}$ \\
      \bottomrule  
\end{tabular}
\caption{Clifford corrections for generalized teleportation of $T$ gate.}
\label{tab:cc}
\end{table}

\emph{Fault tolerance and cost analysis.} 
Fault-tolerance of our scheme follows from fault-tolerance of each part, i.e. 
achieving a target infidelity $\epsilon$ of the gate to be teleported requires the infidelities of in the base code, magic state preparation and Pauli measurements keep at least at the same order.
Consequently, characterized by the total number of qubits needed to execute one logical non-Clifford gate, the cost of our scheme is simply the summation of contributions from different parts,
\begin{equation}
    Q_{\text{total}}(\epsilon) = Q_{\text{base}}(\epsilon) + Q_{\text{magic}}(\epsilon) + Q_{\text{meas}}(\epsilon).
    \label{eq:total overhead}
\end{equation}
Here $Q_{\text{base}}(\epsilon)$ is the number of qubits needed to run the main circuit, including the cost of maintaining the logical qubit in the base code, and potentially the additional cost of performing the measurement-conditioned corrections;
$Q_{\text{magic}}(\epsilon)$ characterizes the cost of preparing the magic resource,
which is typically $\operatorname{polylog} \epsilon$ (or constant with a constant-rate code M) since we achieve the target fidelity by transversal gates in a sufficiently large code M; $Q_{\text{meas}}(\epsilon)$ characterizes the cost of implementing the logical Pauli measurements, for instance the  $M_{\overline{ZZ}}$ and $M_{\overline{X}}$ in Fig.~\ref{fig:circuit} for $T$ gate, which also has a poly-logarithmic scaling in several generalized lattice surgery schemes \cite{Cohen_2022, cross_2024, ide_2024, swaroop_2024}. 

Now, we would like to stress a key observation that the total cost characterized by Eq.~(\ref{eq:total overhead}) has a simple additive structure rather than a multiplicative one, due to the absence of concatenation in our scheme. Recall that in the conventional MSD, the distillation circuit is run on the qubits encoded in base code and therefore resulting a cost which is roughly a product of the ``$Q_{\text{base}}(\epsilon)$'' and ``$Q_{\text{magic}}(\epsilon)$''(more precisely this should be $Q_{\text{distil}}(\epsilon)$ for the MSD, which will be defined more explicitly later).

\begin{figure}
    \centering
    \includegraphics[width=\linewidth]{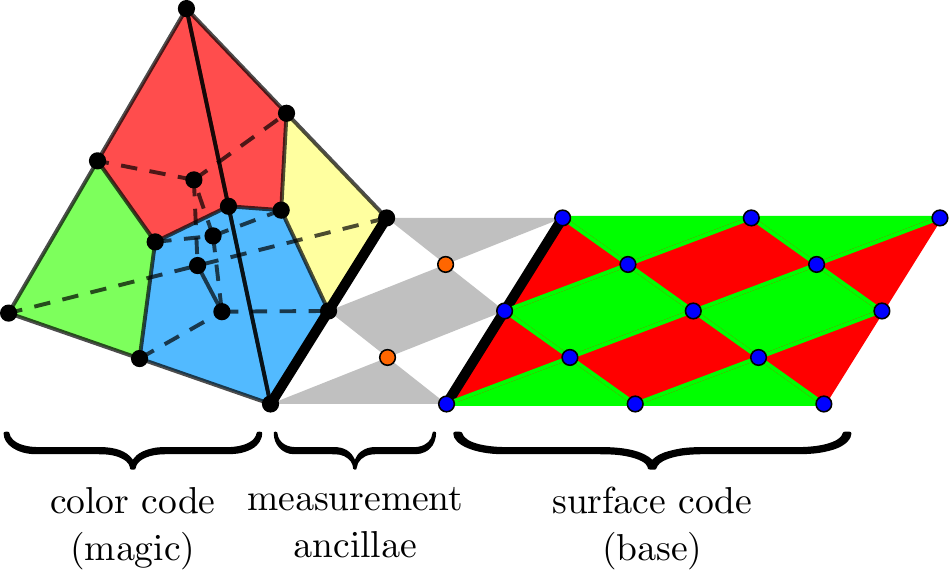}
    \caption{
    Measuring logical $\overline{Z}\overline{Z}$ operator between a distance-3 3D color code (tetrahedron) on the left and a distance-3 surface code (square) on the right, via lattice surgery assisted with 2 ancillary qubits (two dots) in the middle.    %
    The color code (tetrahedron) we use in this protocol consists of 15 qubits represented as black dots at the corners (4), centers of edges (6), faces (4) and body (1). There are in total 4 $X$-checks, corresponding to the hexahedrons at corners. $Z$-checks are quadrilaterals. There are in total 18 of them, with 10 being independent. 
    The logical $\overline{Z}$ we choose in the measurement is supported on the right bottom edge of the tetrahedron, marked by thick line. 
    On the right we put a surface code.
    Blue dots represent 13 physical qubits, 
    solid red diamonds or triangles represent $X$-checks,
    and solid green diamonds or triangles represent $Z$-checks.
    Three qubits on the left boundary form a $\overline{Z}$ operator, marked by thick line. 
    To measure the product $\overline{Z}\overline{Z}$,
    2 qubits represented by red dots are introduced. Then the two 
    $X$-checks (hexahedrons) at the right edge of the color code are enlarged to include the ancillae. Similarly, the two $X$-checks of the surface code on the boundary are enlarged from triangles to diamonds. 
    Finally, 3 extra $Z$-checks are added, represented by grey triangles and diamond,
    whose product 
    is the product of two logical $\overline{Z}$ operators that we want to measure.}
    \label{fig:cc-sc}
\end{figure}

\emph{Examples.}
Here we demonstrate how to implement a logical $T$ gate on a distance-3 surface code, by a generalized gate teleportation from the smallest 3D color code, also known as the $\llbracket 15,1,3\rrbracket$ quantum Reed-Muller (QRM) code (see Fig.~\ref{fig:cc-sc}).
\begin{enumerate}
    \item Prepare a logical $\ket{\overline +}$ in the color code by measuring all $Z$-checks for all physical qubits initialized at the state $\ket{+}$ followed by error correction \cite{Dennis_2002};
    \item 
     Apply transversal logical $\overline{T}$ to obtain the magic state $|\overline{M}\rangle= \overline{T}\ket{\overline{+}}$
    \item Logical joint $\overline{Z} \overline{Z}$ measurement between the two codes, as demonstrated in Fig.~\ref{fig:cc-sc} (a), 
    where 
   2 ancillary qubits and 3 additional $Z$-checks are augmented. 
   Furthermore, the $X$-checks on the both boundaries are extended to cover the ancillary qubits. 
The product of three newly added $Z$-checks give the desired measurement $\overline{ZZ}$.
    \item Decouple the two added ancillary qubits by measuring physical $X$ on each of them.
    \item Logical $\overline{X}$ measurement on the color code. This can be done destructively. 
    For instance, one can measure all physical qubits in $X$-basis and readout the value of a logical $\overline{X}$ operator.
    \item Apply Clifford corrections conditioned on the measurement outcomes $m_{ZZ}, m_X$, following Tab.~\ref{tab:cc}.
\end{enumerate}
One can explicitly check that the merged code keeps the code distance invariant.
Therefore, given physical error rate $p$, we can achieve a logical error rate of order $p^2$ through error detection and post-selection.

The above example can be scaled up to larger codes, i.e. a distance-$d=2l+1$, $l\in \mathbb{Z}$ 3D color codes equipped with transversal $T$ gate \cite{bombin_2015_gauge, Kubica_2015_universal} joint with a distance-$d=2l+1$ surface code. 
In both codes, we can identify the logical $\overline{Z}$ operator with a product of $Z$ operators along a line containing $2l+1$ qubits. Moreover, the $X$-checks, when restrict on this line, resemble the structure for a repetition code. For instance, in Fig.~\ref{fig:cc-sc} the $X$-checks of both $d=3$ ($l=1$) 3D color code and 2D surface code turn to $X$-checks that are product of two nearby qubits when restrict to the thick lines that support the logical $\overline{Z}$. 
For general $l>1$, the repetition code structure for the two logical $\overline{Z}$ from two sides persists, and consequently the lattice surgery employed to perform the joint $\overline{ZZ}$ measurement naturally extends to $l>1$ case. Namely, we supply $2l$ extra qubits as measurement ancillae, extends the $2l$ $X$-checks of both sides to include the ancillae, and add $2l+1$ new $Z$-checks, whose product is exactly the logical $\overline{ZZ}$ operator \footnote{See Supplemental Material B, C at URL-will-be-inserted-by-publisher for more details of the 3D color code and the fault-tolerance of the lattice surgery.}.

Regarding the total overhead, to achieve a target infidelity $\epsilon$, we need $Q_{\text{magic}}(\epsilon) = \Theta(d^3) = \Theta(\log^3\epsilon^{-1})$ physical qubits for magic state preparation,
$Q_{\text{meas},ZZ}(\epsilon)=\Theta(d) = \Theta(\log \epsilon^{-1})$ qubits for the logical $\overline{Z}\overline{Z}$ measurement. 
Logical $\overline{X}$ of the 3D color code can be measured destructively, requiring no more ancillary qubits. 
Finally, Clifford corrections can be done without introducing more physical qubits utilizing long-range connectivity in hardware. 
To summarize, the overall overhead of this scheme is dominated by the cost of preparing the magic state using a 3D color code, leading to $Q_{\text{total}}=\Theta(\log^3\epsilon^{-1})$.
As we will see later, this is more efficient than the $\Theta(\log^{4.46}\epsilon^{-1})$ total overhead from the conventional MSD with $\llbracket 15, 1, 3\rrbracket$ 3D color code as the distillation circuit running on a fixed-size surface code.

The above $\log^3 \epsilon^{-1}$ scaling can be improved by taking M from a family of qLDPC codes with high rate and transversal non-Clifford gates
\cite{lin_2024, golowich_2024_ldpc, breuckmann_2024}. For instance, let us assume that code M has parameters scaled as $\llbracket n, k=\Omega(n^{1-\alpha}), d=\Omega(n^\beta)\rrbracket$.
Then, the overhead for magic state preparation is $Q_{\text{magic}}(\epsilon) = (n/k)|_{d=\Theta(\log\epsilon^{-1})}=\Theta(\log^{\alpha/\beta} \epsilon^{-1})$. 
On the other hand, state-of-the-art measurement protocol can reach an overhead of $Q_{\text{meas}} = \mathcal{O}(d\log^2 d)$ for codes M and B with distance $d = \Theta(\log \epsilon^{-1})$ \cite{swaroop_2024}. Therefore, 
as long as $\alpha/\beta < 1$, the total overhead is dominated by the cost of Pauli measurement, which is $\mathcal{O}(\log \epsilon^{-1} \log^2 \log \epsilon^{-1})$.

\emph{Comparison with the conventional magic state distillation.} 
At high level, our scheme uses a different strategy to achieve magic state at arbitrary precision. More explicitly, to prepare a magic state with a target fidelity, 
in the conventional MSD schemes, one picks a fixed code and iterates a fixed-size sub-program for sufficient times,
while in our scheme one picks a sufficiently large code from a family of codes.
As a result, the MSD scheme always has a finite break-even point (sometimes called a ``distillation threshold''),  while our scheme requires the code family to have a finite threshold. 
In the following, we will compare the overhead of the MSD schemes and ours, assuming the existence of code families with the required properties.
We denote the code for MSD as D(istillation) with parameters $\llbracket n_{\text{D}}, k_{\text{D}}, d_{\text{D}}\rrbracket$, and denote the code for our scheme as M.

For MSD with small codes, such as the $\llbracket 15, 1, 3\rrbracket$ 3D color code, multiple rounds of distillation are performed, utilizing error detection and post-selection in each round to reduce the error in output magic states. 
In this case, the initial number of noisy magic ancillary qubits needed for distilling a magic state with error rate $\epsilon$ is $Q_{\text{distil}}=\log^\gamma \epsilon^{-1}$, where $\gamma = \log (n_{\text{D}}/ k_{\text{D}}) / \log d_{\text{D}}$ \cite{Bravyi_2012}. 
Since MSD should be performed at the logical level, the total number of qubits for producing one magic state is schematically
\begin{equation}
    Q_{\text{MSD}}(\epsilon) = Q_{\text{base}}(\epsilon)\times Q_{\text{distil}}(\epsilon).
    \label{eq:overhead-small-msd}
\end{equation}
For example, MSD based on the $\llbracket 15, 1, 3\rrbracket$ 3D color code has $\gamma \approx 2.46$.~\footnote{
Recently, a single-round MSD scheme is proposed to achieve constant-overhead MSD \cite{wills_2024}.
In this scheme, a sufficiently large code D is chosen from a family of codes
such that concatenating D with the base code B, one round of distillation is sufficient to yield a magic state in code B with the target error rate $\epsilon$.
The cost of such a distillation scheme is still given by Eq.~(\ref{eq:overhead-small-msd}) due to the concatenation structure,
with $Q_{\text{distil}}(\epsilon)$ being $(n_{\text{D}}/k_{\text{D}})|_{d_{\text{D}}=\Theta(\log\epsilon^{-1})}$,
directly comparable to $Q_{\text{magic}}(\epsilon)$ in our scheme.
Therefore, our scheme turns multiplication in the  distillation-teleportation scheme to addition. }
Accompanied with a surface code as base code for the logical processor, the overall overhead scales as $\log^{4.46}\epsilon^{-1}$, larger than the $\log^3 \epsilon^{-1}$ overhead of color code-surface code teleportation scheme.

The comparison above is made without any optimization for both schemes, serving as a conceptual demonstration of the potential advantage of our distillation-free scheme.
It is also interesting to compare the optimal performance for both schemes
with surface code chosen as the base code.
MSD scheme can then be optimized by using large base code only in the late rounds,
since surface codes can be expanded fault-tolerantly \cite{Dennis_2002}.
Consequently, the cost of MSD is dominated by the final round of distillation 
\footnote{A careful analysis shows that if $n_{\mathrm{D}}<k_{\mathrm{D}}d_{\mathrm{D}}^2$, then the cost is dominated by the final round of distillation and thus scales as $\log^2\epsilon^{-1}$, 
which is the case for many distillation protocols 
\cite{Campbell_2017,haahCodesProtocolsDistilling2018}.
Otherwise the cost is dominated by the initial noisy ancillae and thus scales as $\log^\gamma \epsilon^{-1}$. 
The parameter $\llbracket 15, 1, 3\rrbracket$ falls into this category.
Since we are discussing the optimal performance, we focus on the first case here.
Note that further optimization can be adopted to lower the number of qubits needed, but it does not change the scaling $c\log^2\epsilon^{-1}$ \cite{haahCodesProtocolsDistilling2018}.
}.
Therefore, the number of additional qubits (not including qubits for running the main circuit) required by the optimized MSD scheme scales as $c\log^{2}\epsilon^{-1}$ with $c>1$ an $\epsilon$-independent factor determined by code D. 
On the other hand, with code M taking from suitable qLDPC codes as discussed in the \textit{Examples}, the additional cost of our scheme is dominated by the cost of logical Pauli measurements, which scales as $\log \epsilon^{-1} \log^2\log\epsilon^{-1}$ in the state-of-the-art measurement scheme \cite{swaroop_2024}.
In short, with surface code chosen as the base code, the MSD scheme requires at least one surface code ancilla for implementing one logical non-Clifford gate, while the number of additional qubits required by our scheme can be lower in scaling.

\emph{Summary and discussion.} 
We propose a distillation-free scheme for implementing logical non-Clifford gates based on a main circuit with fault-tolerant Clifford operations. 
Our scheme is based on a generalized gate teleportation protocol to teleport the non-Clifford gate executed on magic ancillae to the main circuit, such that the magic ancillae and the main circuit are allowed to be protected by different codes.
This scheme leverages advancements in large codes with transversal non-Clifford gates and developments in lattice surgery techniques.
We compare our scheme with the distillation-teleportation scheme, focusing on the requirement of the codes and the scaling of the overhead required to achieve a target fidelity.

In addition to the potential advantage in asymptotic overhead, we also want to comment on the differences between two schemes from a practical point of view. 
Firstly, the input of a distillation procedure consists of a large number of encoded magic states. 
Preparing such encoded magic states below the distillation threshold has been a challenging task, as encoding a general quantum state into a quantum CSS code is far less straightforward than preparing the code in the computational basis \cite{Dennis_2002,Li_2015, Chamberland_2019, Chamberland_2020, Lao_2022, daguerre_2024,Ye_2023, Gupta_2024, kim_2024}. 
To input such states, a gate teleportation subroutine is required. 
In contrast, our scheme directly takes physical non-Clifford gates as its input, which is conceptually more straightforward. 
Secondly, the MSD requires unitary encoding or decoding at logical level to prepare the magic resource, or even deeper circuits under the space-time tradeoff \cite{haahCodesProtocolsDistilling2018}. 
Clock time for such logical operations is determined by the code cycle of the base code B.
In contrast, our scheme prepares the magic state through a transversal gate followed by error correction, which is one code cycle of the code M, providing a potential to reduce the time cost in preparing the magic ancilla.

Our scheme shares similarity with code switching proposals for logical non-Clifford gates \cite{Anderson_2014, Kubica_2015_universal, Bombín_2016, Vasmer_2019} in that they both use code deformation techniques.
However, in a code switching scheme, the physical qubits of the base code B with transversal Clifford gates (such as the 2D color code) form a subset of physical qubits of the code M with a transversal non-Clifford gate (such as the 3D color code). This subsystem structure necessitates executing the non-Clifford gate after switching from B to M and before switching back, which makes such schemes difficult to modularize. 
In contrast, our scheme allows for an off-line magic factory for creating and maintaining fault-tolerant magic ancillae, as it is similar to the distillation-based scheme in that they both separate the production of magic states and gate teleportation.

Finally, we comment on the experimental feasibility of our scheme based on the current technology: (1) the beyond-2D connectivity that is generally required to generate magic resource has be realized in existing platforms such as trapped ion and atom array \cite{moses_2023,Bluvstein_2023, Wang_2024}; 
(2) the gate teleportation in our scheme is performed by logical measurement, whereas the generalized lattice surgery techniques reduce it to the task of merely measuring the stabilizers of a qLDPC code \cite{Cohen_2022} -- a routine in maintaining a qLDPC quantum memory and therefore does not introduce additional experimental challenges. 

\emph{Acknowledgment}
We thank Min-Hsiu Hsieh, Alexei Kitaev and Zi-Wen Liu 
for discussions. 
This work is supported by the National Key R\&D Program of China 2023YFA1406702, Tsinghua University Dushi program and DAMO Academy Young Fellow program.

\bibliography{draft}

\end{document}


\title{Supplemental Material: Magic teleportation with generalized lattice surgery}

\author{Yifei Wang}
\author{Yingfei Gu}
 \email{guyingfei@tsinghua.edu.cn}
\affiliation{Institute for Advanced Study, Tsinghua University, Beijing 100084, China}

\date{\today}

\maketitle

\onecolumngrid

\appendix

\onecolumngrid

\section{A. Generalized gate teleportation for diagonal gates $U$
}

In this section we discuss generalized gate teleportation for an $n$-qubit diagonal gate, and give explicit instructions for Clifford corrections associated with the diagonal gates in 3rd Clifford hierarchy, i.e. $T$, C$S$ or CC$Z$.

Let $\overline{U}$ be a diagonal gate in computational basis, with the following standard form 
\begin{equation}
    \overline{U}\ket{\overline{x}} = \exp(\mathrm{i}\pi f(x))\ket{\overline{x}},\quad x \in \mathbb{F}_2^n, \quad f:\mathbb{F}_2^n \rightarrow \mathbb{R}/2\mathbb{Z}
\end{equation}
The overline in expressions such as $\ket{\overline{x}}$
emphasizes the logical-level object, for the application in our scheme. 
For gates in 3rd Clifford hierarchy, with $n=1$ and $f(x) = x/4$, $U$ is the $T$ gate, 
with $n=2$ and $f(x_1,x_2) = x_1x_2/2$, $U$ is the C$S$ gate,
and with $n=3$ and $f(x_1,x_2,x_3) = x_1x_2x_3$, $U$ is the CC$Z$ gate.

\begin{figure}
    \centering
    \includegraphics[width=.6\linewidth]{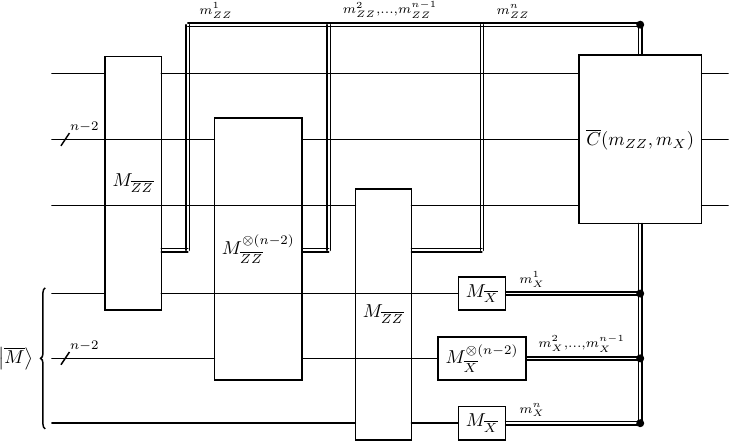}
    \caption{Gate teleportation for an $n$-qubit diagonal gate.
    We collapse qubits 2 to $n-1$ into one line, which represents 
    $n-2$ folds of same operations as those for qubits 1 and $n$.}
    \label{fig:gt-gen}
\end{figure}

The circuit for gate teleportation is shown in Fig.~\ref{fig:gt-gen}.
It includes (logical) $\overline{ZZ}$ measurements across different codes,
$\overline{X}$ measurements on the magic factory (also referred as ancillary qubits in this section),
and the correction $\overline{C}(m_{ZZ},m_X)$ conditioned on the measurement outcomes $m_{ZZ},m_X$, that is applied to the main circuit.

The magic state $\ket{\overline{M}}$ for gate $\overline{U}$ is
\begin{equation}
    \ket{\overline{M}} = \overline{U}\ket{\overline{+}}^n = 2^{-n/2}\sum_{y\in \mathbb{F}_2^n}\mathrm{e}^{\mathrm{i}\pi f(y)}\ket{\overline{y}}.
\end{equation}
Consider computational basis $\ket{\overline{x}}$ $(x\in\mathbb{F}_2^n)$ for the main circuit.
After measurements of $ZZ$ and then $X$, which give outcomes $m_{ZZ}\in \mathbb{F}_2^n$ and $m_{X}\in\mathbb{F}_2^n$ respectively, the system evolves to 
\begin{equation}
\begin{aligned}
    \ket{\overline{x}}\ket{\overline{M}} & \xmapsto{M_{ZZ}} \mathrm{e}^{\mathrm{i}\pi f(x\oplus m_{ZZ})}\ket{\overline{x}}\ket{\overline{x\oplus m_{ZZ}}}\\ 
    &\xmapsto{M_X} \mathrm{e}^{\mathrm{i}\pi (f(x\oplus m_{ZZ})+(x\oplus m_{ZZ})\cdot m_X)}\ket{\overline{x}}\overline{H}^{\otimes n}\ket{\overline{m_X}},
\end{aligned}
\end{equation}
where $\cdot$ denotes the standard inner product in the vector space $\mathbb{F}_2^n$.
At this moment, the state of the ancillary qubits depends only on the measurement outcome and is independent of $x$, showing that these qubits have been disentangled from the main circuit and can be safely discarded, which also implies that the measurement of $M_{\overline{X}}$ can be done destructively.  

The remaining task is to append the correct phase $\exp(\mathrm{i}\pi f(x))$ to the state $\ket{x}$, which necessitates further corrections on the main circuit. 
Explicitly, we need to append a correction phase $\mathrm{i}\pi c(x)$ to the basis state $\ket{\overline{x}}$, where
\begin{equation}
    c(x) = f(x) - f(x\oplus m_{ZZ})-(x\oplus m_{ZZ})\cdot m_X.
\end{equation}
The phase $-\mathrm{i}\pi m_{ZZ} \cdot m_X$ is an overall phase independent of $x$. 
The phase $-\exp(\mathrm{i}\pi x\cdot m_X)$ can be appended by applying Pauli $\overline{Z}$ operators on the main circuit, conditioned on the measurement outcome $m_X$. 
Finally, note that
\begin{equation}
    \mathrm{e}^{\mathrm{i}\pi (f(x)-f(x\oplus m_{ZZ}))}\ket{\overline{x}} = \overline{X}(m_{ZZ})\overline{U}^\dag \overline{X}(m_{ZZ}) \overline{U}\ket{\overline{x}},
\end{equation}
where $X(y)$ for $y\in\mathbb{F}_2^n$ is defined to be $ \prod_{i=1}^n X_i^{y_i}$, i.e. for each non-zero element in $y$, one need to apply $X$ to the corresponding qubit.
Therefore, the correction operator is given by 
\begin{equation}
    \overline{C}(m_{ZZ},m_X) = \overline{X}(m_{ZZ})\overline{U}^\dag \overline{X}(m_{ZZ}) \overline{U} \overline{Z}(m_X).
    \label{eq: general correction}
\end{equation}
If $U$ lies in the $k$-th Clifford hierarchy,
$C(m_{ZZ},m_X)$ is a diagonal gate in the $(k-1)$-th Clifford hierarchy 
due to the definition and properties of Clifford hierarchies \cite{Gottesman_1999}
and that all diagonal gates in a Clifford hierarchy form a group \cite{Cui_2017}.

As a special case that $U$ is in the 3rd Clifford hierarchy, which can be $T$, C$S$ or CC$Z$, only Clifford corrections on the main circuit are needed for its teleportation.
It can be seen from Eq.~(\ref{eq: general correction}) that
the Clifford correction is a product of mutually commuting diagonal gates,
each conditioned on some (combination of) measurement outcomes,
which we call the `key' to the gate.
Therefore, by explicitly carrying out Eq.~(\ref{eq: general correction}),
we list the gates in Clifford corrections for teleporting C$S$ and CC$Z$ and their keys
in Tabs.~\ref{tab:cc-cs}.

\begin{table}
\centering
    \begin{tabular}{cc}
    \multicolumn{2}{c}{C$S$ gate} \\
    \toprule  
   key & gate in $C(m_{ZZ},m_X)$\\
   \midrule  
   $m_{X,1}+m_{ZZ,1}m_{ZZ,2}$ & $\overline{Z}_1$ \\
   $m_{X,2}+m_{ZZ,1}m_{ZZ,2}$ & $\overline{Z}_2$ \\
   $m_{ZZ,2}$ & $\overline{S}_{1}^\dag$ \\
   $m_{ZZ,1}$ & $\overline{S}_{2}^\dag$ \\
    $m_{ZZ,1}+m_{ZZ,2}$ & $\overline{\mathrm{C}Z}_{1,2}$ \\
      \bottomrule  
\end{tabular}
\hspace{30pt}
\begin{tabular}{cc}
\multicolumn{2}{c}{CC$Z$ gate} \\
    \toprule  
   key & factor in $C(m_{ZZ},m_X)$\\
   \midrule  
   $m_{X,1}+m_{ZZ,2}m_{ZZ,3}$ & $\overline{Z}_1$ \\
   $m_{X,2}+m_{ZZ,3}m_{ZZ,1}$ & $\overline{Z}_2$ \\
   $m_{X,3}+m_{ZZ,1}m_{ZZ,2}$ & $\overline{Z}_3$ \\
   $m_{ZZ,1}$ & $\overline{\mathrm{C}Z}_{2,3}$ \\
   $m_{ZZ,2}$ & $\overline{\mathrm{C}Z}_{3,1}$ \\
   $m_{ZZ,3}$ & $\overline{\mathrm{C}Z}_{1,2}$ \\
      \bottomrule  
\end{tabular}
\caption{Clifford corrections for generalized teleportation of C$S$ gate (left) and CC$Z$ gate (right).
For example, when conducting an C$S$ gate teleportation,
if one gets measurement outcome $m_{ZZ}=(1,1)$, $m_X = (0,1)$, 
then gates corresponding to keys $m_{X,1}+m_{ZZ,1}m_{ZZ,2}$,
$m_{ZZ,1}$, $m_{ZZ,2}$ should be applied,
i.e., $\overline{C} = \overline{S}_1\overline{S}^\dag_2$.
}
\label{tab:cc-cs}
\end{table}

\section{B. An explicit construction of 3D color codes}

In this section we present an explicit construction of 3D color codes, which is equivalent to that provided in \cite{bombin_2015_gauge} and used in \cite{Beverland_2021_cost}. 
We will describe its parameters and the relevant logical $\overline{Z}$ operator we are to measure. 

A color code is generally defined on a colorable simplicial complex.
In particular, a 3D color code $C$ is defined on a 4-colorable 3-dimensional simplicial complex $\mathcal{L}$ such that 
each vertex $v$ of the complex is assigned to a color among a color set $\{\text{r}, \text{g},\text{b},\text{y}\}$, and 
there is no edge that connects two vertices of the same color. 
The boundary of $\mathcal{L}$, denoted by $\partial\mathcal{L}$,
is defined in the sense of a subset of $\mathbb{R}^3$.
Then, the qubits and checks of $C$ are given as follows
\begin{itemize}
    \item Qubits are assigned to all 3-simplices (tetrahedrons); 
    \item $X$-checks are assigned to all 0-simplices (vertices) except those in $\partial\mathcal{L}$ 
    \item $Z$-check are assigned to all 1-simplices (edges) except those in $\partial\mathcal{L}$ 
\end{itemize}
Here, when we assign checks to 0-simplices or 1-simplices, we are checking all the qubits (3-simplices) that contain the corresponding  0-simplices or 1-simplices. Namely, the stabilizer $S_\sigma^{X}$ or $S_\sigma^{Z}$ on a simplex $\sigma$ is given as 
\begin{equation}
    S^X_\sigma = \prod_{\delta_3\supset \sigma} X_{\delta_3},  \quad  S^Z_\sigma = \prod_{\delta_3 \supset \sigma} Z_{\delta_3}, 
\end{equation}
where $\delta_3$ represents 3-simplices that host qubits. 
The commutation relations between $X$- and $Z$-checks are ensured by the property
of colorable complexes \cite{Kubica_2015_universal}.

We now show an explicit construction of a series of 
 4-colorable simplicial complexes $\mathcal{L}_l$ indexed by an integer $l$, which corresponds to a family of 3D color codes with increasing distance $d=2l+1$. 

\begin{figure}
    \centering
    \includegraphics[width=.7\linewidth]{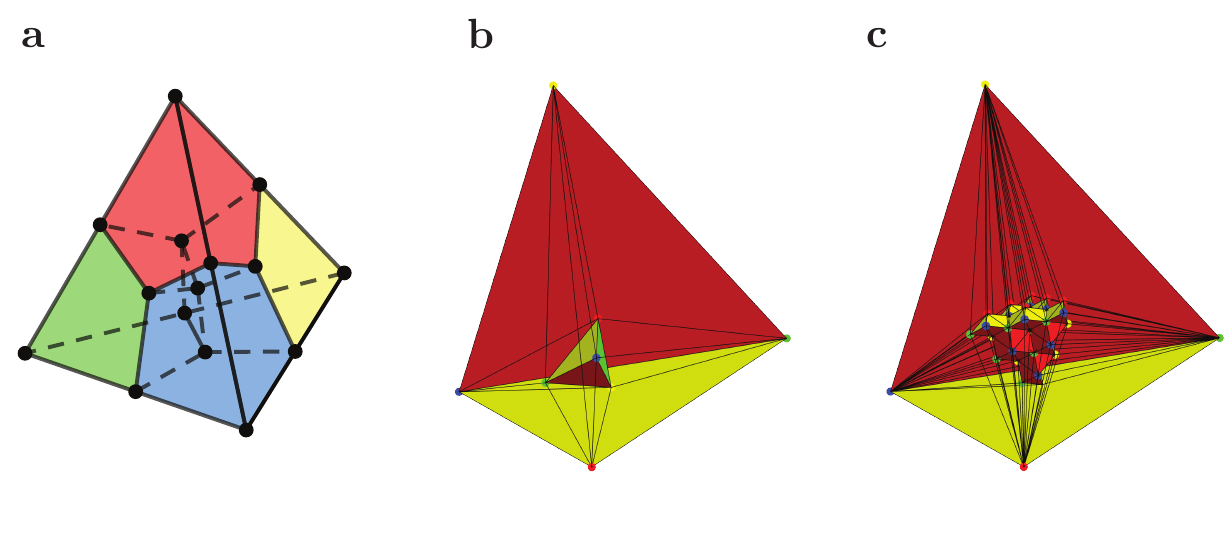}
    \caption{The smallest 3D color code in original picture (a) and dual picture (b). (c) is for general code in this series with code distance $2l+1$. 
Note only two faces of the outer tetrahedron in (b, c) are colored for visual convenience.
    }
    \label{fig: minimal 3dcc}
\end{figure}

Let us start with $l=1$, for which we will show its equivalence to the $d=3$ tetrahedron color code on a dual lattice that is used in the main text.
The construction involves two tetrahedrons, see Fig.~\ref{fig: minimal 3dcc} (b). We refer them as the inner tetrahedron and the outer tetrahedron in this section. 
Both tetrahedron consist of 4 vertices in 4 different colors.
The 4 vertices, 6 edges and 4 faces of the outer tetrahedron form the boundary $\partial\mathcal{L}_1$. 
Each vertex of the outer tetrahedron is put towards the boundary face of the inner tetrahedron with the same color, 
and edges are added between the outer vertex and the vertices on the boundary face (if the vertex has color r, then the boundary face of the inner tetrahedron towards it has color r, and the vertices on the face have colors g, b or y).
All individual triangles and tetrahedrons are identified as 2- and 3-simplices. At this point, following the rules that assign qubits to 3-simplices, $X$-checks to 0-simplices and 
$Z$-checks to 1-simplices except those on the $\partial \mathcal{L}_1$, we already obtain a 3D color code with $15$ qubits, 4 $X$-checks and 18 (10 of which are independent) $Z$-checks. 

Now, we further show that the above construction from $\mathcal{L}_1$  is dual to
the one used in the main text.
The inner tetrahedron in (b) is dual to the body center in (a).
The tetrahedron formed by the inner r-vertex and the outer r-face (or other 3)
in Fig.~\ref{fig: minimal 3dcc} (b) is dual to the 
vertex in the r-corner (or other 3) in Fig.~\ref{fig: minimal 3dcc} (a).
The tetrahedron formed by the inner rb-edge 
and the the outer gy-edge (or other 5) in (b)
is dual to the middle point of the edge
between r- and b-corners (or other 5) in (a).
The tetrahedron formed by the inner y-face 
and the outer y-vertex (or other 3)
is dual to the center of face crossing
r-, g- and b-corners (or other 3) in (a).
The 4 inner vertices in (b), representing 4 $X$-checks,
is dual to 4 corner hexahedrons of the same color in (a).
The 18 edges that are not in the boundary in (b),
representing $Z$-checks,
is dual to 18 faces in (a).
Actually, we can also view (b) in a mixed view,
only referring to the inner tetrahedron.
That is, we represent the simplices connecting inner and outer tetrahedrons
by their intersection with the inner tetrahedron.
This automatically reduces these tetrahedrons to vertices, edges and faces in the boundary of the inner tetrahedron, 
leading to a more direct connection to (a).
Furthermore, reducing edges connecting an inner vertex
and many outer vertices to the inner vertex just removes some redundancy in the $Z$-checks and does not change the code.

\begin{figure*}
    \centering
    \includegraphics[width=\linewidth]{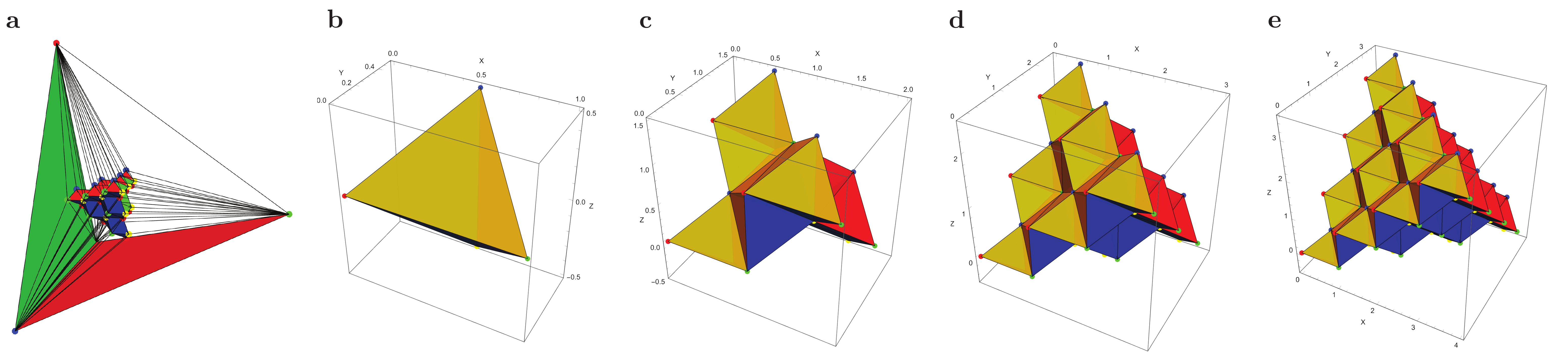}
    \caption{Illustration of lattices $\mathcal{L}_l$. (a) Lattice $\mathcal{L}_3$.
Generally, complex $\mathcal{L}_l$ is composed of an inner lattice being a large tetrahedron formed by many 3-simplices, and an outer tetrahedron serving as the boundary.
The inner lattice is carved out of a bcc lattice whose vertices have coordinates
$(x,y,z)$ and $(x,y,z)+(1/2,1/2,1/2)$ in $\mathbb{R}^3$, $x,y,z\in\mathbb{Z}$.
(b-e) Inner lattices of $\mathcal{L}_1$ to $\mathcal{L}_4$.
To build the inner lattice of $\mathcal{L}_l$, we first choose an original tetrahedron with vertices 
$O_{\text{r}}=(0,0,0)$,
$O_{\text{g}}=(1,0,0)$,
$O_{\text{b}}=(1/2,1/2,1/2)$,
$O_{\text{y}}=(1/2,1/2,-1/2)$,
each assigned to the color represented by its subscript.
We further choose three basis vectors 
$\hat{i}=(0,1,1)$,
$\hat{j}=(1,0,1)$,
$\hat{k}=(1,1,0)$,
and let the vertices with color $c\in\{\text{r}, \text{g},\text{b},\text{y}\}$ have coordinates
$O_{c}+x\hat{i}+y\hat{j}+z\hat{k}$, where $x,y,z\in \mathbb{Z}$
with $x,y,z\geq 0$ and $x+y+z<l$.
A 1-simplex (edge) of the complex is a set containing 2 vertices whose Euclidean distance is no more than 1.
A 2-simplex (face) of the complex is a set containing 3 edges which form a triangle, along with the 3 vertices of the triangle.
A 3-simplex (tetrahedron) of the complex is a set containing 4 faces which form a tetrahedron, along with the edges and vertices of the tetrahedron.
A face composing vertices of 3 different colors (say r, g, b) are assigned to the one rest color among the four colors (say y).
For showing the structure of the lattice, the b- and y-faces of the outer tetrahedron and the faces connecting inner and outer lattices are not plotted in color in (a).
    }
    \label{fig: lattice of 3dcc}
\end{figure*}

Next, we construct the simplicial complex $\mathcal{L}_l$  for $l>1$ by replacing the inner tetrahedron structure by a large lattice shown in Fig.~\ref{fig: minimal 3dcc} (c). 
The inner tetrahedron is carved out of a body-centered cubic (bcc) lattice.
See Fig.~\ref{fig: lattice of 3dcc} for details.

The upshot is that, we get a 4-colored 3-dimensional simplicial complex $\mathcal{L}_l$ containing
$(2l(l+1)(l+2)/3+4)$ 0-simplices,
$(14l^3/3+8l^2+16l/3+6)$ 1-simplices,
$(8l^3+12l^2+8l+4)$ 2-simplices and
$(4l^3+6l^2+4l+1)$ 3-simplices.
As for the color code $C_l$ defined on the lattice $\mathcal{L}_l$,
the number of qubits equals to the number of 3-simplices,
\begin{equation}
    n = 4l^3 + 6l^2 + 4l+1.
\end{equation}
The number of $X$-checks equals to the number of inner vertices,
\begin{equation}
    r_X = \frac{2}{3}l^3 + 2l^2 + \frac{4}{3}l.
\end{equation}
Since each vertex poses two restrictions on the $Z$-checks \cite{Kubica_2015_universal}, the number of $Z$-checks equals to the number of inner edges minus twice the total number of vertices, which is
\begin{equation}
    r_Z = \frac{10}{3}l^3 + 4l^2 + \frac{8}{3}l.
\end{equation}
The number of logical qubits is $n-r_X-r_Z = 1$ as expected.
The logical Pauli operators of the code constructed as above is given by strings or membranes of some shrunk complex \cite{Bombin_2007_exact}.
As a consequence, the distance of code $C_l$ is $(2l+1)$.

The logical $\overline{Z}$ operator that is contained in the product $\overline{Z}\overline{Z}$ we want to measure is a
minimal weight operator supported on an edge of the lattice $\mathcal{L}_l$.
For example, consider the edge linking the r- and b- vertices in the outer tetrahedron (the leftmost edge in Fig.~\ref{fig: lattice of 3dcc} (a)).
This edge is in $(2l+1)$ 3-simplices of $\mathcal{L}_l$,
corresponding to $(2l+1)$ qubits supporting the logical $Z$ operator.
These qubits are involved in $2l$ $X$-checks, which are represented by 
y- and g-vertices between the b- and r-boundary faces of the inner tetrahedron.
In the mixed view in which simplices linking to the boundary tetrahedron are reduced to their intersections with the inner tetrahedron, 
the support of this $\overline{Z}$ operator is simply the y-g boundary edge of the inner tetrahedron (bottom-right edges in Fig.~\ref{fig: lattice of 3dcc} (b-e)), 
where qubits lie on the $(2l-1)$ 1-simplices and 2 endpoints,
checked by $2l$ $X$-checks represented by all vertices on this edge.
Viewing qubits as classical bits, and $X$-checks as parity checks, the $(2l+1)$ bits and $2l$ checks form a classical repetition code.

\section{C. Measuring the $ZZ$ operator between color code and surface code}

In this section, we demonstrate how to measure the logical $\overline{Z}\overline{Z}$
operator between a color code and a surface code by merging and splitting the codes.
We prove that the distance of the merged code is no less than that of the original color code and surface code, which implies fault-tolerance of the measurement.

From the previous section we see that there is a logical $\overline{Z}$ operator 
supported by $(2l+1)$ qubits with $2l$ $X$-checks forming a repetition code.
We want to measure the product of this logical $\overline{Z}$ with 
a logical $\overline{Z}$ operator in a distance-$(2l+1)$ surface code
supported by a smooth boundary.
Since in the surface code,
the $X$-checks involving the qubits on the smooth boundary
are represented by the vertices on the boundary,
the checks and qubits also form a repetition code as in the color code.

\begin{figure}
    \centering
    \includegraphics[width=.4\linewidth]{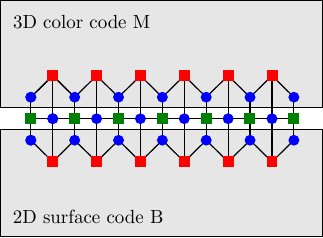}
    \caption{Illustration for measuring the logical $\overline{Z}\overline{Z}$ 
    operator between a color code and a surface code, each with distance $(2l+1)$, $l=3$.}
    \label{fig:cc-tc-gen}
\end{figure}

As is shown in Fig.~\ref{fig:cc-tc-gen}, to measure the product of these two operators, we introduce $2l$ ancillary qubits (blue circles) between these two boundary lines, each put between a pair of $X$-checks (red squares). 
Each of the $2l$ $X$-checks in the color code is extended to involve one ancillary qubit, and so does each of the $2l$ $X$-checks in the surface code.
We further introduce $(2l+1)$ $Z$ checks (green squares), each involves a qubit from the color code, a qubit from the surface code, and one or two newly added ancillary qubits.
It is easy to check that the new $Z$-checks we introduce commute with the extended $X$-checks, and that
the product of these $(2l+1)$ newly added $Z$ checks is exactly the operator $\overline{Z}\overline{Z}$ we want to measure. 

To show that this construction is fault-tolerant, 
we need to prove that the merged code has a distance no less than $(2l+1)$.
Let's first consider the $X$-distance.
Since the operator $\overline{ZZ}$ is turned into a stabilizer,
the single-qubit logical $\overline{X}$ can no longer be a valid logical operator.
The only $X$-type logical operator is now $\overline{XX}$,
which is the product of logical $\overline{X}$ operators in the color and surface code before merging.
After merging, the only kind of distortion we can do to a logical $\overline{XX}$ operator
is multiplication with some $X$-checks in the original codes.
If the $X$-check we multiply is inside the codes,
it will certainly not decrease the $X$-distance.
If the $X$-check we multiply is extended to the measurement ancillae, 
it will only increase the weight of a minimal-weight $X$-type logical operator,
and hence not decrease the distance.

The maintenance of $Z$-distance is slightly more non-trivial.
Pictorially, the logical $\overline{Z}$ operator in the merged code can only be 
(i) a logical $\overline{Z}$ operator in the surface code, multiplied by some $Z$-checks of the color code and some newly added $Z$-checks, or
(ii) a logical $\overline{Z}$ operator in the color code, multiplied by some $Z$-checks of the surface code and some newly added $Z$-checks.
For case (i), multiplying stabilizer can only affect weight of a logical $Z$ operator that is supported by some qubits on the boundary, that is, qubits appearing in Fig.~\ref{fig:cc-tc-gen}.
It is trivial that multiplying only $Z$-checks in the color code can only increase the weight.
We first show that multiplying newly added $Z$-checks does not decrease the weight.
In fact, each of an added $Z$-check `flips' a Pauli $Z$ supported on the surface code boundary to a Pauli $Z$ supported on the color code boundary, which cannot reduce the weight of a $Z$ operator.
Finally, we show that multiplying $Z$-checks in the color code after flipping some Pauli $Z$ operators from the surface code boundary to the color code boundary does not reduce the weight.
In fact, the whole color code boundary $B$ supports a logical $Z$ operator of minimal weight.
Suppose we have partly flipped the $\overline{Z}$ operator in surface code to a subset $F$ of $B$. The color code checks we are to multiply is supported by $S$. Now the qubits in the color code that supports the $\overline{Z}$ operator is $F\cup S \backslash (F\cap S)$. Note that
\begin{equation}
    \begin{aligned}
        |F\cup S \backslash (F\cap S)| &= |F| + |S| - 2 |F\cap S| \\ 
        & \geq |F| + |S| - 2 |B\cap S| \\ 
        & = |F| + (|B| + |S| - 2|B\cap S| - |B|) \\
        & \geq |F|.
    \end{aligned}
\end{equation}
The second line holds since $F\subseteq B$. The last line holds since B supports a minimal weight logical operator in the color code.
Case (ii) holds from the same argument with color code and surface code interchanged.
Therefore, we have completed the proof that our construction of the merged code does not reduce the code distance.

Finally, we would like to comment that 
the structure of the $X$- and $Z$-type logical operators discussed pictorially above
can be characterized more formally in an algebraic way,
by the scheme of homological measurement \cite{ide_2024}.
Actually, when doing a CSS type logical measurement, say measuring a $Z$-type logical operator,
the maintenance of $X$-distance is ensured by the scheme itself,
while the maintenance of $Z$-distance should be discussed case by case.

\bibliography{supp}